# Aging and scaling of aftershocks

Sumiyoshi Abe[1] and Norikazu Suzuki[2]

[1]*Institute of Physics, University of Tsukuba, Ibaraki 305-8571, Japan*

[2]*College of Science and Technology, Nihon University, Chiba 274-8501, Japan*

A part of the seismic time series, in which the Omori law for temporal pattern of aftershocks holds, is refereed to as the Omori regime. Here the properties of correlation between earthquake events both inside and outside of the Omori regime are studied by analysis of the data taken in southern California. It is found that inside of the Omori regime correlation exhibits the aging phenomenon, in marked contrast to the fact that no aging is observed outside of the Omori regime. The scaling nature of aging is also discussed and the form of the function responsible for it is presented. The results indicate that complex fault-fault interactions may be governed by glassy dynamics.

A large shallow earthquake tends to reorganize the stress distribution in the relevant area that leads to the swarm of aftershocks. A complex energy landscape with a number of local minima is created, the fault-fault interactions may be long-range and random, and shortness of the event time intervals implies the quenched disorder. The (modified form of) Omori law [1,2] states that the number of aftershocks, $dN(t)$, occurring between the short time interval $t \sim t+dt$, where $t$ stands for time elapsed after the mainshock, obeys

$$\frac{dN(t)}{dt} \sim \frac{1}{(c+t)^p}, \tag{1}$$

where $c$ is a positive constant and the exponent, $p$, empirically ranges from about 0.8 to 1.5. The Omori regime, in which this law holds, plays a distinguished role in the seismic time series because of the *slow* power-law decay of the rate of frequency, $dN(t)/dt$, and nonstationarity. In this Letter, we report the discovery of the aging phenomenon and the scaling law for aftershocks. By contrast, no aging is observed outside the Omori regime. These results and observations indicate that the complex fault-fault interactions for aftershocks may be governed by glassy dynamics [3,4].

Our idea is as follows. We propose to employ as the basic physical random variable the time of the $n$th aftershock with arbitrary magnitude, $t_n$, following the mainshock at $t_0$. Then, we define the correlation function for the aftershock events as

$$C(n+n_w, n_w) = \frac{<t_{n+n_w} t_{n_w}> - <t_{n+n_w}><t_{n_w}>}{\sqrt{\sigma^2_{n+n_w} \sigma^2_{n_w}}}, \qquad (2)$$

where the averages and the variance are given by $<t_m> = (1/N) \sum_{k=0}^{N-1} t_{m+k}$, $<t_m t_{m'}> = (1/N) \sum_{k=0}^{N-1} t_{m+k} t_{m'+k}$, and $\sigma^2_m = <t^2_m> - <t_m>^2$, respectively. It clearly satisfies the initial condition: $C(n_w, n_w) = 1$. If the series $\{t_m\}$ is "stationary", then the correlation function in Eq. (2) becomes independent of the "waiting time", $n_w$, whereas "nonstationarity" is characterized by the $n_w$ dependence of the correlation function.

We have analyzed the seismic data provided by the Southern California Earthquake Data Center (available at http://www.scecdc.scec.org/catalogs.html). Among extensive data analyses, we here present two typical examples. Their mainshocks are M7.3 at 11:57:34.1 on June 28, 1992 (34°12.01'N latitude, 116°26.20'W longitude, and 0.97 km in depth) and M7.1 at 9:46:44.13 on October 16, 1999 (34°35.63'N latitude, 116°16.24'W longitude, and 0.02 km in depth). According to the data between 1984 and 2002, the maximum depth is 57.88 km. Therefore, these mainshocks are considered to be very shallow, having in fact been followed by the swarms of aftershocks. For these mainshocks, we have identified the corresponding Omori regimes by examining the best fits of the (modified form of) Omori law in Eq. (1) to the data based on the method of least squares. (However, it does not seem to be necessary to present such identifications here.) An important point is that we do not put restrictions on the area for the aftershocks but consider whole of the relevant area, according to the *nonreductionistic* philosophy for complexity of the earthquake phenomenon. Recently, we have studied the statistical properties of the three-dimensional distance between two successive earthquakes [5] and the network structure [6]. We have discovered that the distance distribution obeys Tsallis nonextensive statistics [7,8] and the earthquakes yields an evolving scale-free network [9,10]. These results highlight the complex-system aspect of the earthquake phenomenon.

In Figs. 1 and 2, we present the plots of the correlation functions with respect to the event number, $n$, for various values of the "waiting time", $n_w$, inside of the Omori regimes. The clear aging phenomenon is appreciated. On the other hand, the correlation functions outside of the Omori regime are plotted in Fig. 3. Almost no aging is observed there. This shows a distinguished nonstationary feature of the Omori regimes.

A point of interest regarding aging depicted in Figs. 1 and 2 is that it is possible to realize the data collapses by simply rescaling $n$ by a certain function, $f(n_w)$. By this manipulation, Figs. 1 and 2 are transformed to Figs. 4 and 5, respectively. The value of the function, $f(n_w)$, for each $n_w$ is determined in such a way that the discrepancies between the correlation functions measured by the $l^1$ norm distance becomes minimum. Also, $f(0)=1$ should hold. The corresponding values of $f(n_w)$ obtained by this procedure are shown as the dots in Fig. 6 and 7. It turns out that they are well fitted by a two-parameter function

$$f(n_w) = a n_w^\gamma + 1, \tag{3}$$

with which now the scaling law

$$C(n+n_w, n_w) = \tilde{C}(n/f(n_w)) \qquad (4)$$

may be established.

In conclusion, we have discovered the aging phenomenon and the scaling law of aftershocks inside of the Omori regime through analysis of the seismic data taken in southern California. Quenched disorder, slow relaxation (the power-law nature of the Omori law), the aging phenomenon, and the scaling law are all characteristics in physics of glasses. The present results and these observations indicate that complex fault-fault interactions leading to aftershocks may be governed by glassy dynamics. This opens a new possibility of approaching the yet unclear mechanism of aftershocks by the use of available physical models such as spin glasses.

# Figure Captions

Fig. 1  Plots of the correlation functions with respect to the event number, $n$, inside of the Omori regime of the earthquake M7.3 in 1992, in the dimensionless units. The values of the "waiting time" are, $n_w = 0$, 600, 1200, and 1800. $N = 15000$. The aging phenomenon is clearly appreciated.

Fig. 2  Plots of the correlation functions with respect to the event number, $n$, inside of the Omori regime of the earthquake M7.1 in 1999, in the dimensionless units. The values of the "waiting time" are, $n_w = 0$, 400, 800, and 1200. $N = 15000$. The aging phenomenon is clearly appreciated.

Fig. 3  Plots of the correlation functions with respect to the event number, $n$, outside of the Omori regime of the earthquake M7.1 in 1999, in the dimensionless units. The values of the "waiting time" are, $n_w = 0$, 400, 800, and 1200. $N = 10000$. No aging phenomenon is recognized.

Fig. 4  The data collapse corresponding to Fig. 1 by rescaling of "time" by the factor $f(n_w)$. The associated values of $f(n_w)$ are given in Fig. 6. All quantities are dimensionless.

Fig. 5  The data collapse corresponding to Fig. 2 by rescaling of "time" by the factor $f(n_w)$. The associated values of $f(n_w)$ are given in Fig. 7. All quantities are dimensionless.

Fig. 6  The values of $f(n_w)$ fitted by the function in Eq. (3) with $a = 1.37 \times 10^{-6}$ and $\gamma = 1.62$, corresponding to Fig. 4. Here, more values of $n_w$ are examined than in Fig. 4. All quantities are dimensionless.

Fig. 7  The values of $f(n_w)$ fitted by the function in Eq. (3) with $a = 1.70 \times 10^{-4}$ and $\gamma = 1.00$, corresponding to Fig. 5. Here, more values of $n_w$ are examined than in Fig. 5. All quantities are dimensionless.